\documentclass{aa}
\usepackage{txfonts}
\usepackage{graphicx}

\begin{document}

\title{Signatures of tidal disruption in the 
Milky Way globular cluster NGC\,6981 (M72)}

\author{Andr\'es E. Piatti\inst{1,2}\thanks{\email{andres.piatti@unc.edu.ar}}, Mart\'{\i}n F. Mestre\inst{3,4}, 
Julio A. Carballo-Bello\inst{5}, Daniel D. Carpintero\inst{3,4}, Camila Navarrete\inst{6,7}, Marcelo D. Mora\inst{8}, and Carolina Cenzano\inst{8}}


\institute{Instituto Interdisciplinario de Ciencias B\'asicas (ICB), CONICET-UNCUYO, Padre J. Contreras 1300, M5502JMA, Mendoza, Argentina
\and Consejo Nacional de Investigaciones Cient\'{\i}ficas y T\'ecnicas (CONICET), Godoy Cruz 2290, C1425FQB,  Buenos Aires, Argentina
\and Instituto de Astrof\'{\i}sica de La Plata (CONICET-UNLP), Argentina
\and Facultad de Ciencias Astron\'omicas y Geof\'{\i}sicas de La Plata (UNLP), Argentina
\and  Instituto de Alta Investigaci\'on, Universidad de Tarapac\'a, Casilla 7D, Arica, Chile
\and  European Southern Observatory, Alonso de C\'ordova 3107, Casilla 19001, Santiago, Chile
\and Millenium Institute of Astrophysics, Av. Vicu\~na Mackenna 4860, 782-0436 Macul, Santiago, Chile
\and Instituto de Astrof\'isica, Facultad de F\'isica, Pontificia Universidad Cat\'olica de Chile, Av. Vicu\~na Mackenna 4860, 782-0436 Macul, Santiago, Chile\\
}

\date{Received / Accepted}

\abstract{
We study the outer regions of the Milky Way globular cluster NGC\,6981 from publicly
available $BV$ photometry and new Dark Energy Camera (DECam) observations, both
reaching nearly 4 mag below the cluster main sequence (MS) turnoff. While the $BV$ data sets
reveal the present of extra-tidal features around the cluster, the much larger field of view of
DECam observations allowed us  to identify some other tidal features, which extend from the cluster
toward the opposite direction to the Milky Way center. These 
cluster structural features arise from stellar density maps built using MS stars, once 
the cluster color-magnitude diagram was cleaned from the contamination of field stars.
We also performed $N$-body simulations in order to help us to 
 understand the spatial distribution of the extra-tidal debris. The outcomes reveal the presence
 of long trailing and leading tails mostly parallel to the direction of the
cluster velocity vector. We found that the cluster has lost most of its mass by tidal disruption
during its perigalactic passages, that lasted nearly 20 Myr each. Hence, a decrease in the density 
of escaping stars near the cluster is expected from our $N$-body simulations, which in turn means that
stronger  extra-tidal features could be found out by exploring much larger areas around NGC\,6891.
}


\titlerunning{NGC\,6981}

\authorrunning{A.E. Piatti et al. }

\maketitle

\markboth{A.E. Piatti et al.: }{NGC\,6981}

\section{Introduction}

Globular clusters formed in external galaxies and then gravitationally stripped off  by the 
Milky Way - usually called accreted globular clusters - are expected to show  
evidence of extra-tidal features, like tidal tails,  azimuthally irregular stellar halos, clumpy 
extended structures \citep{carballobelloetal2014,vanderbekeetal2015,kuzmaetal2016,pcb2020}.
During the last few years, deep enough wide-sky photometric surveys have allowed 
to explore the outermost regions of a number of globular clusters, which resulted in the
detection of a variety of extra-tidal structures around them. For instance, tidal tails have been  very recently 
identified in NGC\,362 \citep{carballobello2019}, NGC\,1851, NGC\,2808 \citep{carballobelloetal2018,sollima2020},
4590 \citep{pme2019}, 5139 \citep{ibataetal2019}, 5904 \citep{g2019}, among others. These globular
clusters  are now enlarging the list of nearly 15  globular clusters with detected tidal tails
 \citep[e.g.,  Pal\,5;][]{odenetal2001}. 

\citet{koppelmanetal2019} associated  7 globular clusters to the accreted Helmi streams 
\citep{helmietal1999} based on their kinematic properties: NGC\,4590, 5024, 5053, 5272,
5634, 5904, and 6981.  \citet{massarietal2019} added Pal\,5, Rup\,106, and E\,3
in the candidates list based on their orbital properties and a less restrictive selection criterion.
However, the membership of E\,3 to this group was recently questioned by \citet{forbes2020}. All of them have prograde orbital motions. According to the recent 
classification of globular clusters carried
out by \citet{pcb2020} from studies focused on their outermost regions, we find, among 
those associated to the Helmi streams, 3 globular clusters with observed tidal tails (NGC\,4590,
5904, Pal\,5), 2  with extra-tidal features that are different from tidal tails (NGC\,5053, 5634),
and 3 without any signatures of extended stellar density profiles (NGC\,5024, 5272, Rup\,106). Recently, \citet{carballobelloetal2020} has confirmed the presence of tidal tails emerging from E\,3 using {\it Gaia} DR2 data.
The outermost regions of NGC\,6981  have been studied by 
\citet{grillmairetal1995} from photographic photometry hardly reaching down the main sequence
turnoff \citep[see, also,][]{amigoetal2013}. They estimated a tidal radius of 8.3$\arcmin$  
and found no signature of clear tidal tails.

NGC\,6981 is identified as lying on a sequence of possibly accreted globular clusters in the age-metallicity diagram, which coincides with the Kraken clusters \citep{kruijssenetal2019}.
As can be inferred, the different
proposed progenitors of NGC\,6981 points to the need for further refinement in the different selection procedures \citep{piatti2019}. The ratios of the cluster mass lost by tidal disruption of the
Milky Way gravitation field to the initial cluster mass  ($M_{dis}$/$M_{ini}$) computed by \citet{piattietal2019b} 
result to be relatively small for the 6 Helmi streams globular clusters mentioned by
\citet{koppelmanetal2019} with studies of their outermost regions 
(0.04 $\le$ $M_{dis}$/$M_{ini}$ $\le$ 0.15). Pal\,5 and Rup\,106 have $M_{dis}$/$M_{ini}$ = 
0.24, while NGC\,6981, 0.41.  The $M_{dis}$/$M_{ini}$ ratio was used by
\citet{piattietal2019b} as an indicator for tidal field strength. They studied the relationship 
between $M_{dis}$/$M_{ini}$, the semimajor axes and the eccentricity of the globular clusters'
orbits (see their figure 1), and found that clusters with relatively high $M_{dis}$/$M_{ini}$ are
either clusters with orbit eccentricities $\ga$ 0.7 or semimajor axes $\la$ 3 kpc. Nevertheless,
a puzzling population of clusters with intermediate eccentricities and short semimajor axes is observed, also with relatively high orbital inclinations. 

In this study, we focus on NGC\,6981 with the aim of finding out some trails of tidal tails that
can explain the large amount of mass lost by the interaction with the Milky Way gravitational field.
In section 2 we describe the analysis carried from public data sets, while in Section 3 we
perform a similar analysis from our own more spatially extended observations. Section 4
discusses the present outcomes, while Section 5 summarizes the main conclusions of this work.

\begin{figure}
\includegraphics[width=\columnwidth]{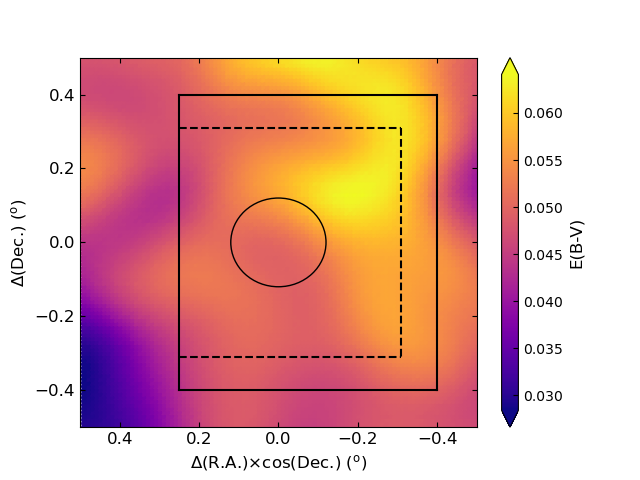}
\caption{Reddening variation across the field of NGC\,6981. The solid box represents the
area of the \citet{stetsonetal2019}'s photometry, while the dashed lines delimite the internal
boundaries of the adopted field star reference field. The circle corresponds to the
cluster tidal radius compiled by \citet{harris1996}.}
\label{fig:fig1}
\end{figure}

\begin{figure}
\includegraphics[width=\columnwidth]{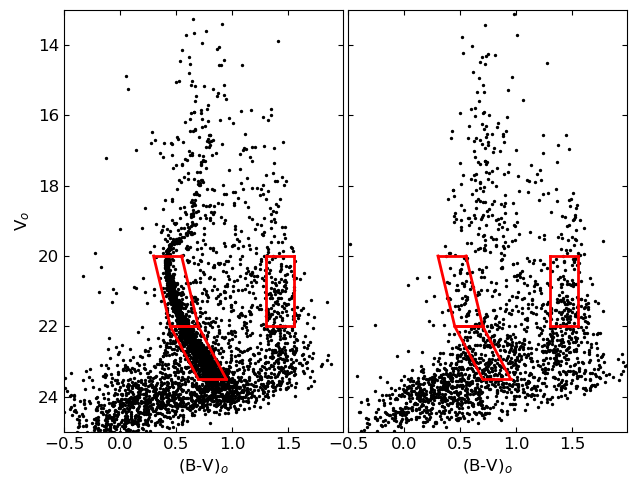}
\caption{Intrinsic CMDs for the cluster field (left panel) for an annulus with internal and external
radii of $0\fdg 05$ and  $0\fdg 12$, respectively  compared to that for a field
region with an equal cluster area (right panel), placed at $\Delta$(Dec.) $> 0\fdg 30$ and 
$\Delta$(RA)$\times$cos(Dec.) $> -0\fdg 12$. Two segments along the cluster MS and
another to redder colors are indicated in red.}
\label{fig:fig2}
\end{figure}

\begin{figure}
\includegraphics[width=\columnwidth]{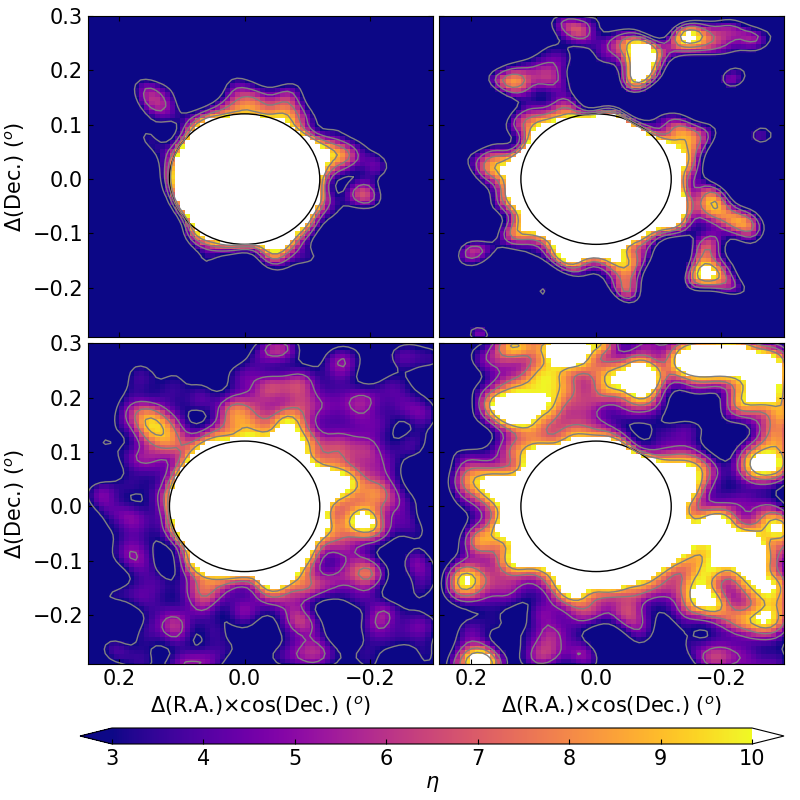}
\caption{Observed (upper panels) and field star cleaned (bottom panels) stellar density maps for
 the brighter (left panels) and the fainter (right panels) cluster MS segments according to Figure~\ref{fig:fig2}. The black circle centered on the cluster indicates the 
cluster tidal radius. Contours for $\eta$ = 3,5,7, and 9 are also shown,  which reflect significance levels.}
\label{fig:fig3}
\end{figure}

\section{Public data handing}

We searched for available unexplored public wide-field photometry to build stellar density maps
from faint cluster main sequence (MS) stars, which are suitable for uncovering extra-tidal 
structures. The homogeneous Johnson $BV$ photometry published by \citet{stetsonetal2019} for 
NGC\,6981, with typical internal and external uncertainties of the order of a few millimagnitudes, 
resulted to be the most appropriate one. The field-of-view is $\sim$ $0\fdg 6 \times 0\fdg 8$,
centered on the cluster, which allows us to examine out to nearly 2 times the cluster tidal
radius \citep[$0\fdg 12$;][2010 Edition]{harris1996}. The reddening variation across the cluster
field is shown in Fig.~\ref{fig:fig1}, with $E(B-V)$ color excess retrieved from \citet{sf11}
provided by the NASA/IPAC Infrared Science Archive\footnote{https://irsa.ipac.caltech.edu/}.
Although the cluster field is not affected by differential reddening, we preferred to employ
dereddened color-magnitude diagrams (CMDs) by correcting the $V$ mags and $B-V$ colors
using the $E(B-V)$ color excesses associated to the positions of the stars in the sky. 
Figure~\ref{fig:fig2} shows the intrinsic CMDs for a cluster region ($0\fdg 05 \le r \le 0\fdg 12$)
compared to a star field region with equal cluster area, located far from the cluster. 

\begin{figure}
\includegraphics[width=\columnwidth]{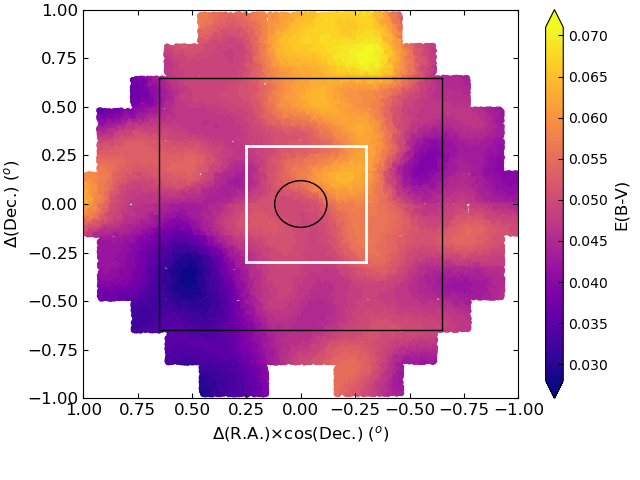}
\caption{Reddening variation across the DECam field of NGC\,6981. The solid white box represents the
area of the \citet{stetsonetal2019}'s photometry (see Fig.~\ref{fig:fig1}), 
while black circle and box represent
the cluster tidal radius compiled by \citet{harris1996}, and the internal
boundary of the adopted field star reference field.}
\label{fig:fig4}
\end{figure}

\begin{figure}
\includegraphics[width=\columnwidth]{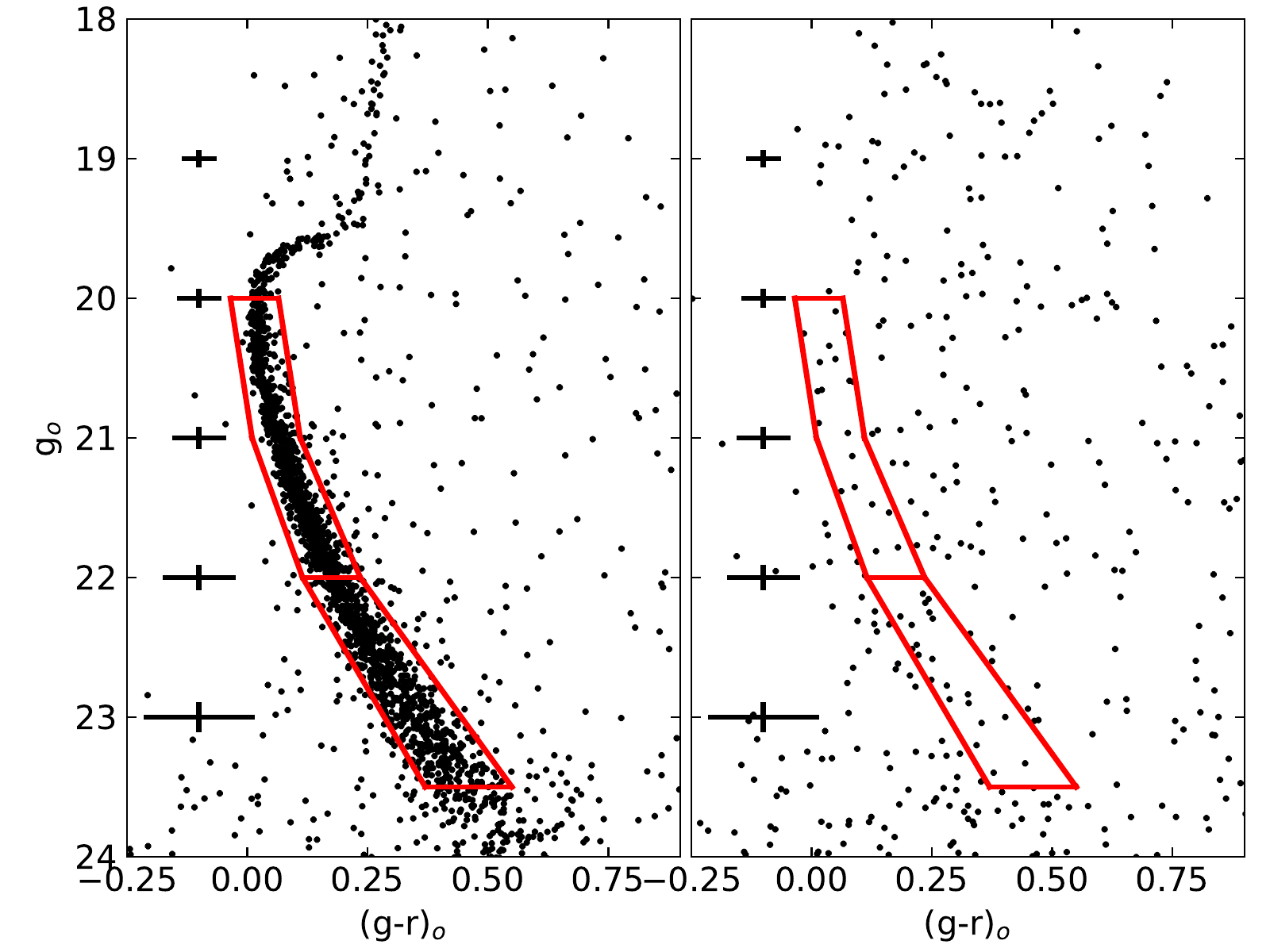}
\caption{Same as Fig.~\ref{fig:fig2} for the present DECam photometry.
Two segments along the cluster MS are indicated with red lines. Error bars are 
also shown.}
\label{fig:fig5}
\end{figure}

\begin{figure}
\includegraphics[width=\columnwidth]{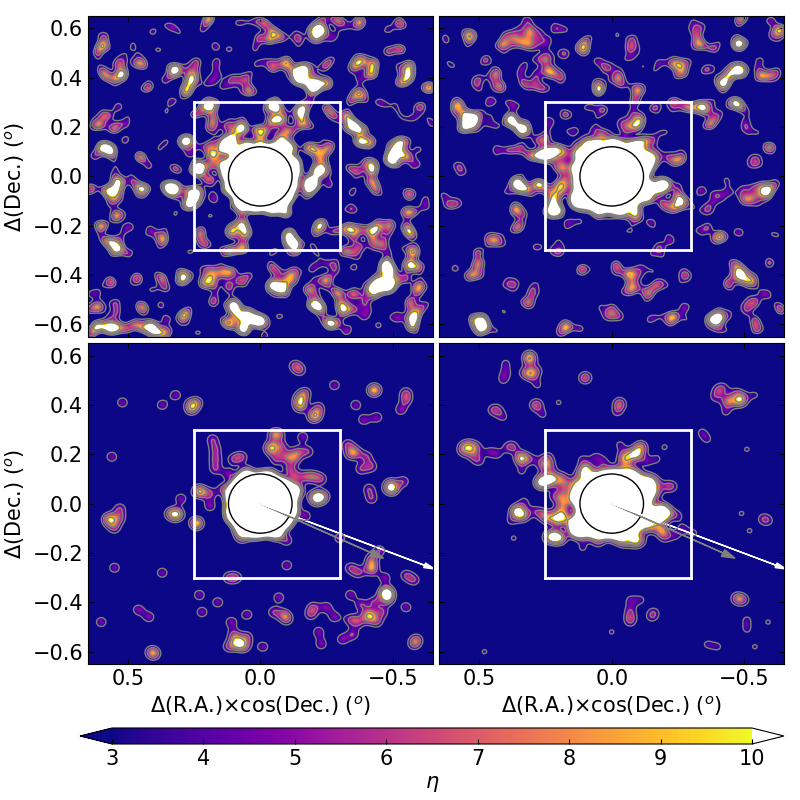}
\caption{Observed (upper panels) and field star cleaned (bottom panels) stellar density maps for
 the brighter (left panels) and the fainter (right panels) cluster MS segments according to Figure~\ref{fig:fig4}. The black circle centered on the cluster indicates the 
cluster tidal radius, while the white box represents the area of the \citet{stetsonetal2019}'s photometry (see Fig.~\ref{fig:fig1}). Contours for $\eta$ = 3,5,7, and 9 are also shown.
Gray and white arrows represent the direction of the motion of the cluster and that
toward the Galactic center, respectively. }
\label{fig:fig6}
\end{figure}

We devised two segments along the cluster MS (see Fig.~\ref{fig:fig2})  where we perform
star counts. Both segments are contaminated by the presence of field stars, so that we first 
applied the procedure proposed by \citet{pb12} to statistically eliminate them.
The method compares the distribution of MS stars within the devised segments, spread across
the cluster field, with that of a reference
star field. For this purpose, we adopted as a reference star field the area delimited by
the dashed lines in Fig.~\ref{fig:fig1} and the data set boundary, where we assumed that
the stellar density, and the distribution in $V$ mag and $B-V$ colors of those stars are
representative of the field stars projected along the line-of-sight of the cluster.
We subtract from each segment a number of stars equal to the corresponding one in the reference 
star field. The distribution of magnitudes  and colors of the 
subtracted stars from the cluster CMD needs to resemble that of the reference star field. 
With the aim of  avoiding stochastic effects caused by very few field stars distributed in less populated CMD regions, we started finding stars to eliminate within a  cell of
($\Delta$$V_0$, $\Delta$$(B-V)_0$) = (1.0 mag,0.5 mag) centered on the magnitude and color
values of each reference field star. Thus, it is highly probable to find a star in the cluster CMD
with ($V_0$,$(B-V)_0$) values within those boundaries around the ($V_0$, $(B-V)_0$) ones
of each field star. In the case that more than one star is located inside that cell, the closest one to the center of that ($V_0$, $(B-V)_0$) cell is subtracted. 
The photometric errors are also taken into account while 
searching for a star to be
subtracted from the cluster CMD. With that purpose, we iterate up to one thousand times 
the comparison between the  ($V_0$,$(B-V)_0$) values  of the reference field star and
those of the stars in the cluster CMD. If a star in the cluster CMD falls
inside the defined cell for the reference field star, we subtract that star. The iterations
were carried out by allowing the  ($V_0$, $(B-V)_0$) values of the star in the cluster CMD 
takes smaller or larger values than the mean ones according to their respective
errors.

Figure~\ref{fig:fig3} (upper panels) depicts the observed stellar density maps built for the two 
MS segments using the \texttt{scikit-learn} software machine learning library \citep{scikit-learn} 
and  its \texttt{gaussian} kernel density estimator (KDE). We employed a grid of 100$\times$100 
boxes onto the cluster area and used a range of values for the
bandwidth from 0.005$\degr$ up to 0.040$\degr$ in steps of 0.005$\degr$ in order to apply the KDE to 
each generated box. We adopted a bandwidth of 0.020$\degr$ as the optimal 
value, as guided by \texttt{scikit-learn}.
We also estimated the background level using the stars distributed within the reference field star
area.  We divided this area in boxes of $0\fdg 10 \times 0\fdg 10$ and counted the number of 
stars inside them. We randomly shifted the boxes by $0\fdg 05$ along 
 the R.A. and Dec. directions and repeated the star counting. Finally, we derived the  mean  value using all the defined boxes.
As for the standard deviation,  we
performed a thousand Monte Carlo realizations using the stars located in the reference field
star area, which were shifted along $\Delta$(R.A.)$\times$cos(Dec.) or $\Delta$(Dec.)
randomly (one different shift for each star) before recomputing the density map. 
The color scale in Fig.~\ref{fig:fig3} represents the absolute deviation from the mean value in the
field in units of the standard deviation, that is,  $\eta$ = (signal $-$ mean value)/standard deviation. 
We have painted white stellar densities with $\eta$ $>$ 10 in order to highlight the least dense structures. 

\begin{figure*}
  \includegraphics[width=0.65\columnwidth]{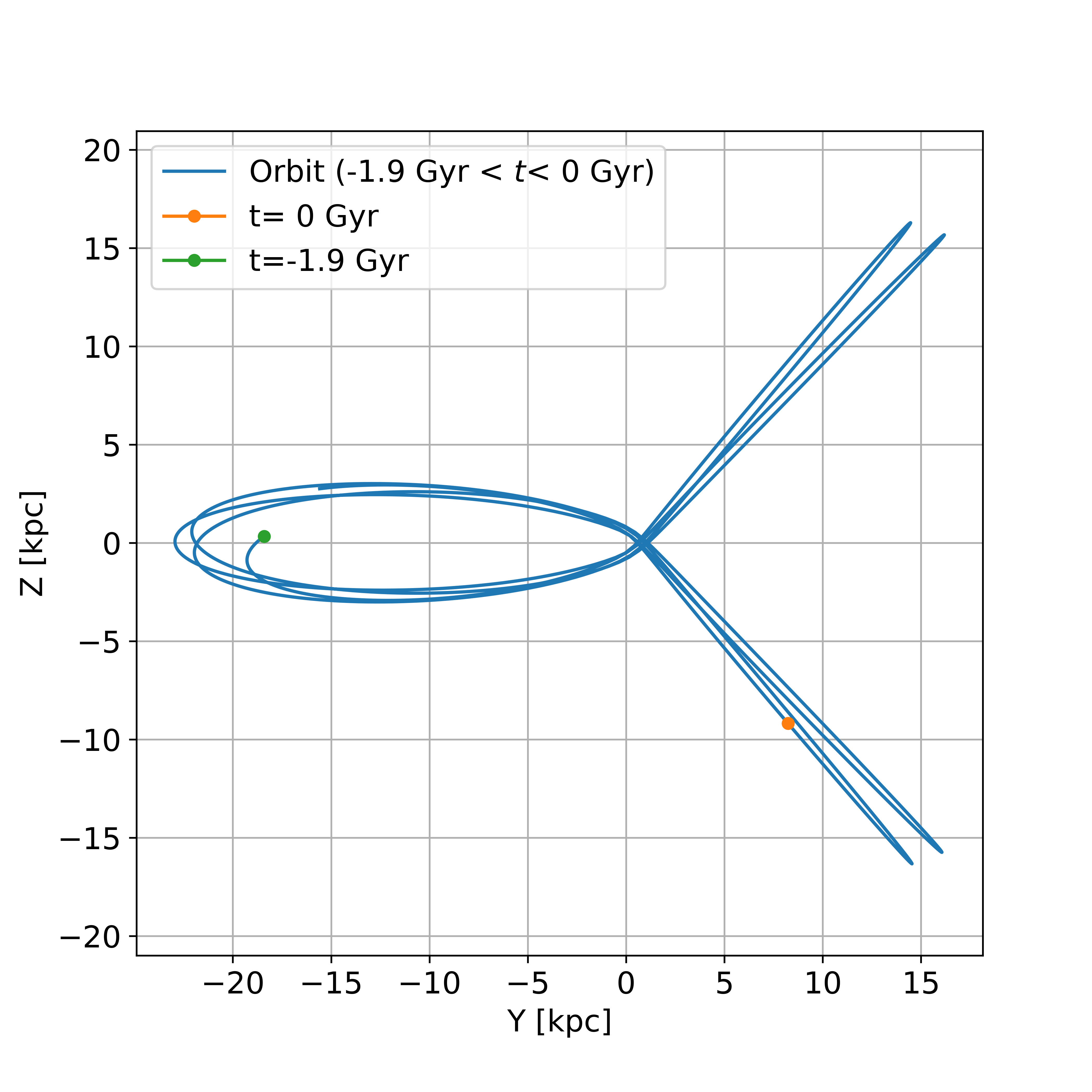}
    \includegraphics[width=0.65\columnwidth]{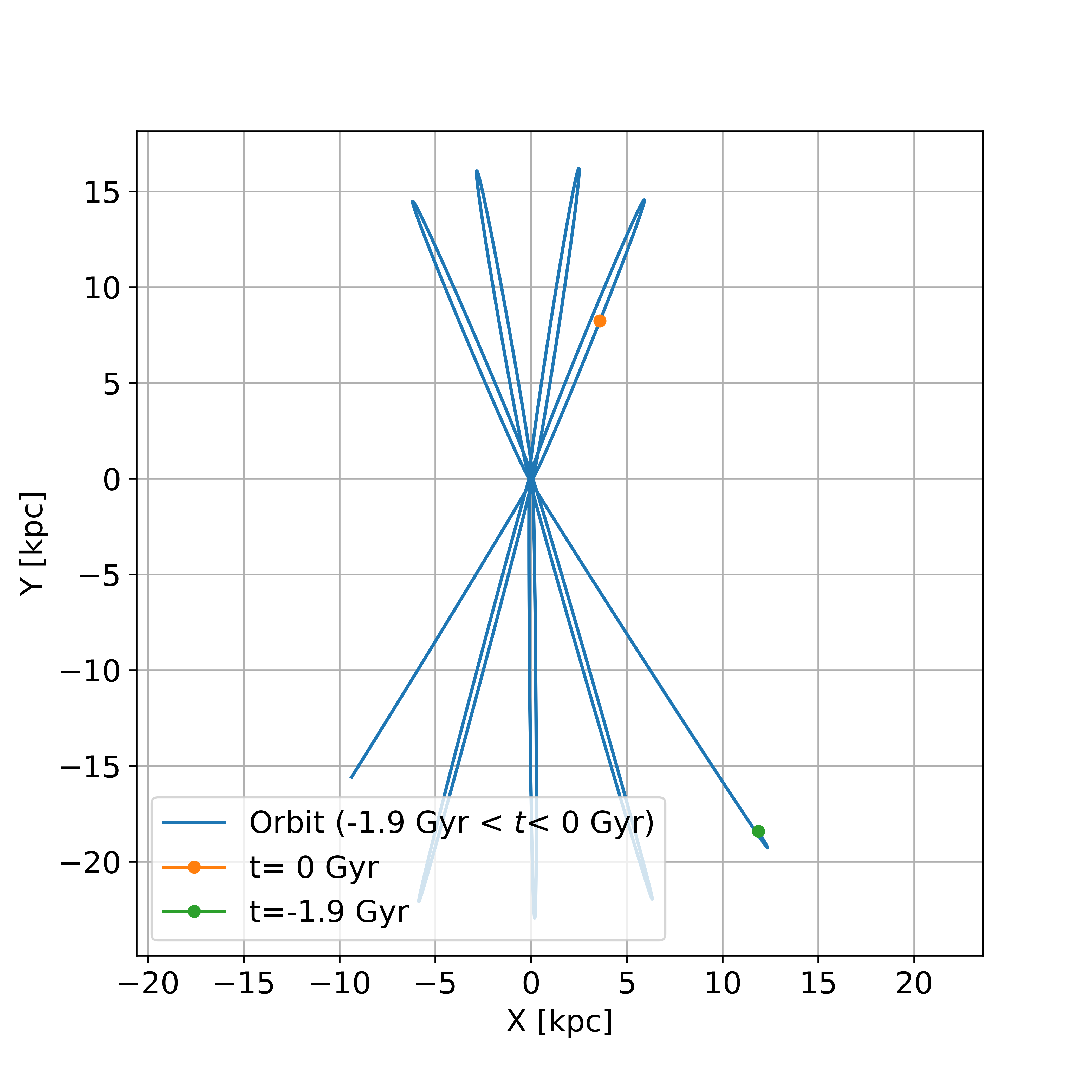}
      \includegraphics[width=0.7\columnwidth]{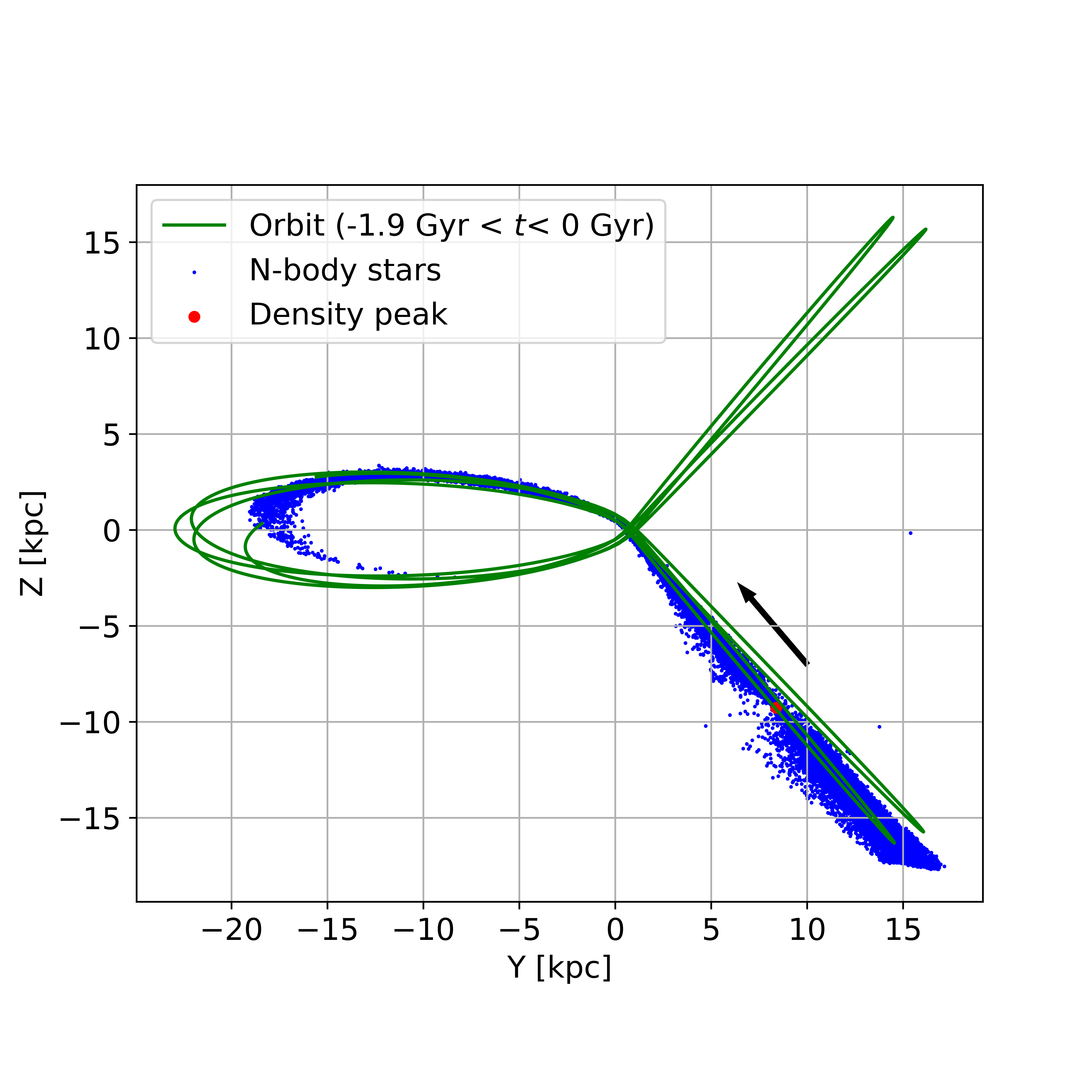}
  \caption{Orbit of the test particle in the $Y-Z$ (left panel) and $X-Y$ (middle panel) planes, 
 respectively.
Present $N$-body snapshot (blue) and orbit of the test particle (green) in the $Y-Z$ plane
(right panel).
  The point of highest density of the $N$-body cluster is marked in red. 
  The black arrow is parallel to the
  velocity of the globular cluster projected in the $Y-Z$ plane.  }
  \label{fig:fig7}
\end{figure*}

We additionally considered the devised segment located to redder colors in the cluster CMD
(the box centered at $(B-V)_0$ $\approx$ 1.5 mag), which represents a composite field star population, and applied the same cleaning procedure.
We found that the resulting cleaned stellar density map does not
contain any visible structure above 1$\eta$, which means that
the residuals of the cleaning technique are negligible. Therefore, we assume that
any stellar enhancement remaining in the stellar density maps  built
from the cleaned cluster CMD is an intrinsic
cluster feature. The resulting field star decontaminated stellar density maps
are shown in the bottom panels of Fig.~\ref{fig:fig3}, where extra-tidal features are readily
visible. As expected, more extended extra-tidal features show up in the case of the 
lowest-mass segment, as lower-mass stars can be more easily stripped away from the cluster 
than their higher-mass counterparts.

\section{Data collection and processing}

With the aim of tracing farther out the cluster extra-tidal features, we carried out observations
(program ID : 2019B-1003)
with the Dark Energy Camera (DECam), attached to the prime focus of the 4-m
 Blanco  telescope at Cerro Tololo Inter-American Observatory (CTIO). DECam provides a 3\,deg$^{2}$ field of view (see Fig.~\ref{fig:fig4}) with its 62 identical chips with a scale of 0.263\,arcsec\,pixel$^{-1}$ \citep{flaugheretal2015}. We observed NGC\,6981 with  the  $g$ and $r$ bands, for which
we obtained 4 exposures of 600 sec. and 400 sec, respectively.
We also observed 3--5 SDSS fields per night at a different airmass to derive the 
atmospheric extinction coefficients  and the transformations between the instrumental magnitudes and the SDSS $ugriz$ system \citep{fukugitaetal1996}.

We processed the images using the DECam Community Pipeline \citep{valdesetal2014},
while the photometry was obtained from the  images with the PSF-fitting
 algorithm of \textsc{daophot\,ii/allstar} \citep{s1987}. The final catalog includes only
  stellar-shaped objects with $|sharpness| \leq  
0.5$ to avoid, as much as possible, the presence of non-stellar sources and background
 galaxies  in our analysis. {\sc daophot\,ii}  was also used to add synthetic stars in our images in order to estimate the completeness of our photometric catalogs. After applying our photometry pipeline on the images with the created artificial stars included, we found 
 that the magnitude for a 50$\%$ recovery of the artificial stars added turned out to be
  23.4\,mag and 23.3\,mag for the $g$ and $r$ bands, respectively.

The methodology that we employed to build the stellar density maps followed the steps 
described in Section 2. We obtained the reddening free $g_0$ magnitudes and $(g-r)_0$
colors by using the individual $E(B-V)$ values (see Fig.~\ref{fig:fig4}) and the
$A_\lambda/A_V$ coefficients given by \citet{wch2019}. Fig.~\ref{fig:fig5} illustrates
the cluster CMD we obtained for the main body of the cluster. Two different segments
were devised along the cluster MS and the field star cleaning procedure was applied.
In this case, we used as the reference star field the region located outside the large black
box drawn in Fig.~\ref{fig:fig4}. Because of the larger area of the DECam FOV with respect to 
\citet{stetsonetal2019}'s FOV, we thoroughly monitor the performance of the cleaning procedure 
by repeating a thousand times the search for a star to be subtracted in the cluster area for each star 
in the reference star field. The position of the subtracted star in the cluster field was chosen
randomly during each simulation of the cleaning procedure. We finally kept those stars that 
remained unsubtracted more than 50$\%$ of the total number of cleaning runs.
The observed and field star cleaned stellar density
maps for both MS segments are depicted in Fig.~\ref{fig:fig6}.

\section{Analysis and discussion}

Fig.~\ref{fig:fig6} shows that NGC\,6981 has visible extra-tidal structures, that can be
described as a non-rounded extended halo and debris distributed along the trailing
tail. \citet{piattietal2019b} computed the Jacobi radii of the cluster for its peri (1.3 kpc) and
apogalactic (26.6 kpc)  distances \citep{baumgardtetal2019}, which resulted to be 
15.5 pc and 67.7 pc, respectively. If
we considered the present cluster Galactocentric distance of 12.85 kpc, its Jacobi radius would be 
$\sim$ 39.0 pc, or $0\fdg 13$
for its heliocentric distance of 17.0 kpc. This Jacobi radius resulted to be similar to the
tidal radius compiled by  \citet{harris1996} and drawn with a black circle in
Fig.~\ref{fig:fig6}. According to the semianalitical model in \citet{piattietal2019b},  who assumed a 50$\%$ cluster
mass loss by evolutionary effects (see their eq. 5 and \citet{baumgardtetal2019}), 
the amount of cluster mass lost
by tidal disruption increases up to 41$\%$ of its initial cluster mass, which means that
the present remaining mass of NGC\,6981 is nearly 9$\%$ of its initial cluster mass.

The stellar density map built by \citet{grillmairetal1995} for
NGC\,6981 using stars distributed in an area of $\sim$ 3$\fdg$3$\times$3$\fdg$3
reveals the presence of some low-level stellar residuals beyond the cluster tidal
radius. Although the authors did not conclude on the existence of tidal tails, their
density map hints for some extra-tidal structure aligned toward the opposite direction
to the Milky Way center. We note that
they used a much shallower photometry than that used in this work.

We performed  $N$-body simulations to investigate the expected tidal features generated 
 only during the last pericentric passages, i.e. the last 2 Gyr of evolution, because
we do not know how long the cluster has been following the present orbit.
We first put a test particle representing the globular cluster at its present 
position and with its present center of mass velocity, and integrated it backwards in time for
a gravitational field model of the Milky Way. Then, we replaced the test particle with an $N$-body system representing the globular cluster, and integrated it forward in time up to the present epoch, embedded in the same host gravitational field. At this point, the tidal features developed by the 
cluster can be compared with those observed.

Following~\citet{baumgardtetal2019}, we modeled the Milky Way with Model I of \citet{irrgangetal2013}, consisting of a bulge, a disk and an halo components.
In order to compute the initial conditions at the present epoch we used \texttt{Astropy} 
\citep{astropy2013,astropy2018} to convert the observed R.A., Dec., distance, radial velocity 
and proper motions given in Table~\ref{tab:tab1} of \citet{baumgardtetal2019} to Galactocentric 
coordinates. We adopted a Cartesian reference frame $(X,Y,Z)$ with corresponding velocities 
$(U,V,W)$ in which the $X$ and $U$ axes point from the Galactic center towards the 
opposite direction of the Sun, $Y$ and $V$ point in the direction of the Galactic rotation at the 
location of the Sun, and $Z$ and $W$ point towards the North Galactic Pole.
Note that this right-handed frame of reference differs from the left-handed one used by \citet{baumgardtetal2019}.  We assumed a distance from the Sun to the Galactic center of $d=8.1$ 
kpc \citep{gravity2018} and a velocity of the Sun respect to the Galactocentric frame of 
$(U,V,W)_\odot=(11.1,252.24,7.25)$ km s$^{-1}$ in agreement with \citet{schonrichetal2010} and 
\citet{rb2004}. The resulting Cartesian coordinates are listed in Table~\ref{tab:tab1}.

\begin{table}
\caption{Present position and velocity of NGC\,6981 in Galactocentric coordinates, used as initial conditions for the backwards ($\Delta t<0$) test particle integration. Positions are given in kpc, 
and velocities in km~s$^{-1}$.
}
\label{tab:tab1}  
\begin{tabular}{@{}lr}\hline\hline
$X_0$ & 3.59746558 \\ 
$Y_0$ & 8.24013064 \\
$Z_0$& -9.17984456\\
$U_0$& -58.75537855  \\
$V_0$&   -147.55728430\\    
$W_0$& 173.06078831 \\
\hline
\end{tabular}
\end{table}

From these initial conditions, we integrated the equations of motion backwards in time during 1.9
 Gyr, an interval chosen in order to start the forward integration near the apocenter of the orbit. Fig.~\ref{fig:fig7} shows the projection on the $Y-Z$ plane of this backward orbit, where we can see a clear $3:2$ resonance (a "fish"). The orbit precesses around the $Z$-axis.

During the $Y>0$ portions of the orbit the cluster moves almost radially (eccentricity $e\leq 1$) deep into the halo with an inclination
$\sim 45\degr$ with respect to the Milky Way plane. During the $Y<0$ portions, the cluster loops perpendicularly to the disk with $e>0$. 
Being the pericentric distance only $\lesssim 1$ kpc, the details of the  integrated orbit may not coincide with the real one, in the sense that  phenomena like e.g. dynamical friction are not taken into account in a test particle integration.

The globular cluster was modeled with an initial King profile \citep{king66}
of total mass $M=2\times 10^5$ M$_\odot$, a King radius $r_0= 2.8$ pc, a King concentration
parameter $c_{\rm{K}}=\log_{10}(r_t/r_0)=1.183$ and a central potential $W_0=5.7$,
using $N=5\times10^4$ stars of 4 solar masses. With these values, other important parameters are:
tidal radius $r_\mathrm{t}=42.64$ pc, core radius $r_\mathrm{c}=2.29$ pc, and concentration parameter
$c=\log_{10}(r_\mathrm{t}/r_\mathrm{c})=1.27$.
This initial parameters are in agreement with those compiled by \citet{harris1996} and \citet{baumgardtetal2019}, which are respectively $c=1.23$ and $r_c=2.33$ pc.
By fitting these parameters at the end of the simulation would have required probing initial values in a trial-and-error approach; therefore, we chose to fix them at the start of the simulation.

The cluster was placed at the final position of
the backward integration and then evolved forward with the code \texttt{Gadget-2}
\citep{springel2005} 
using a maximum step size of $0.01$ Myr and a softening length of $0.5$ pc. We have used a version of the code that allows external potentials to be included in the simulation, in our case that of the Galaxy. Also, the adaptive time-step criterion takes into account the presence of this potential \citep[][Appendix A2.3]{Villalobos_et_Helmi2008}.
Taking into account that the cluster losses mass in this forward integration due to tidal forces, 
the value of the initial mass $M$ was chosen so that the final mass
was close to the observed mass $M_\mathrm{obs}=8.7\times 10^4$ M$_\odot$
\citep{baumgardtetal2019}.
In the simulation the final mass turned out to be $M_\mathrm{f}=9.2\times10^4$ M$_\odot$. 
The difference between the mass loss from the simulation and that of the semianalytical model is expected, given that they are based on different methodologies and hypothesis.
We refer as 'present' to characteristics of this final state.

\begin{figure*}
  \includegraphics[width=0.7\columnwidth]{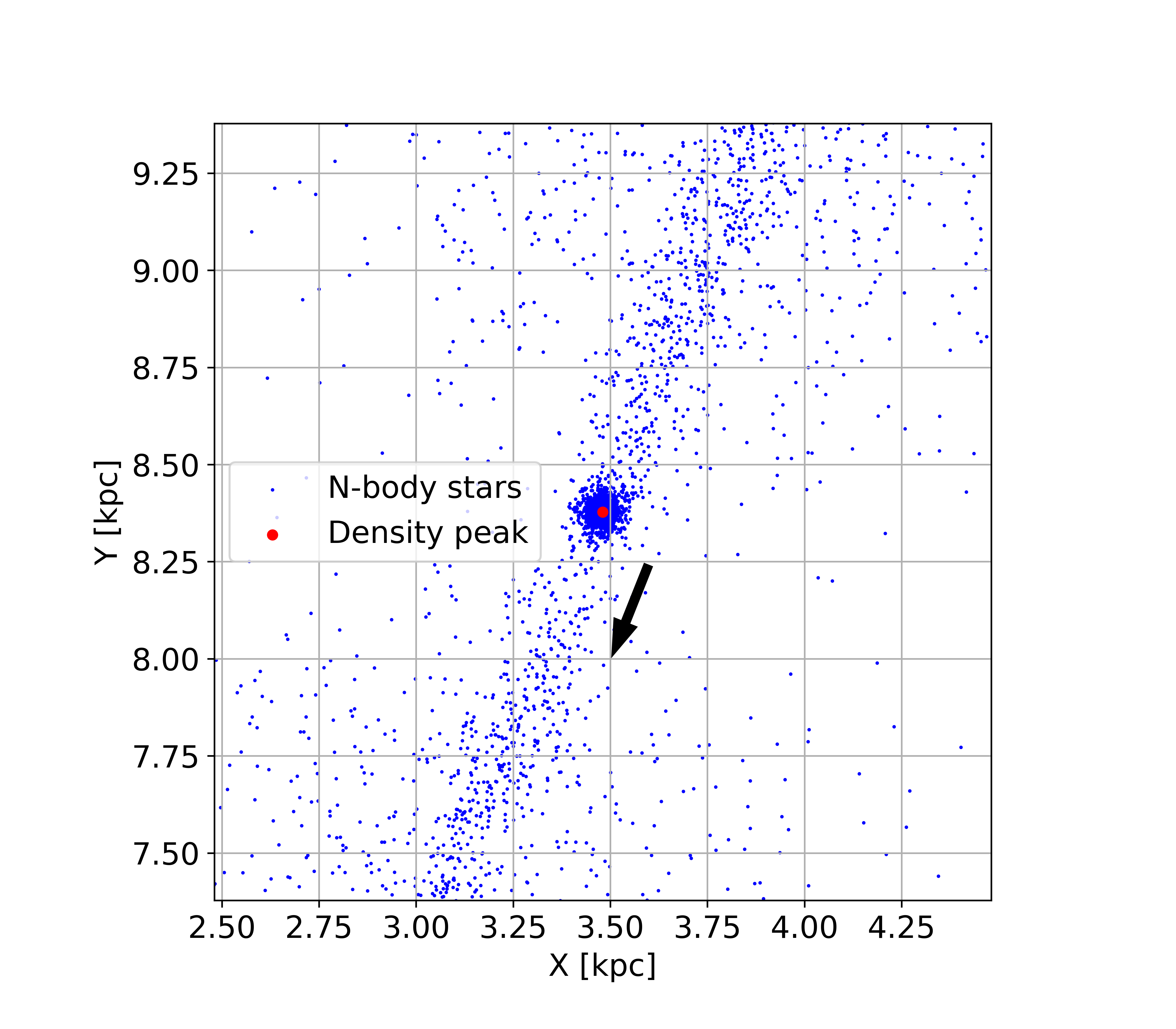}
    \includegraphics[width=0.7\columnwidth]{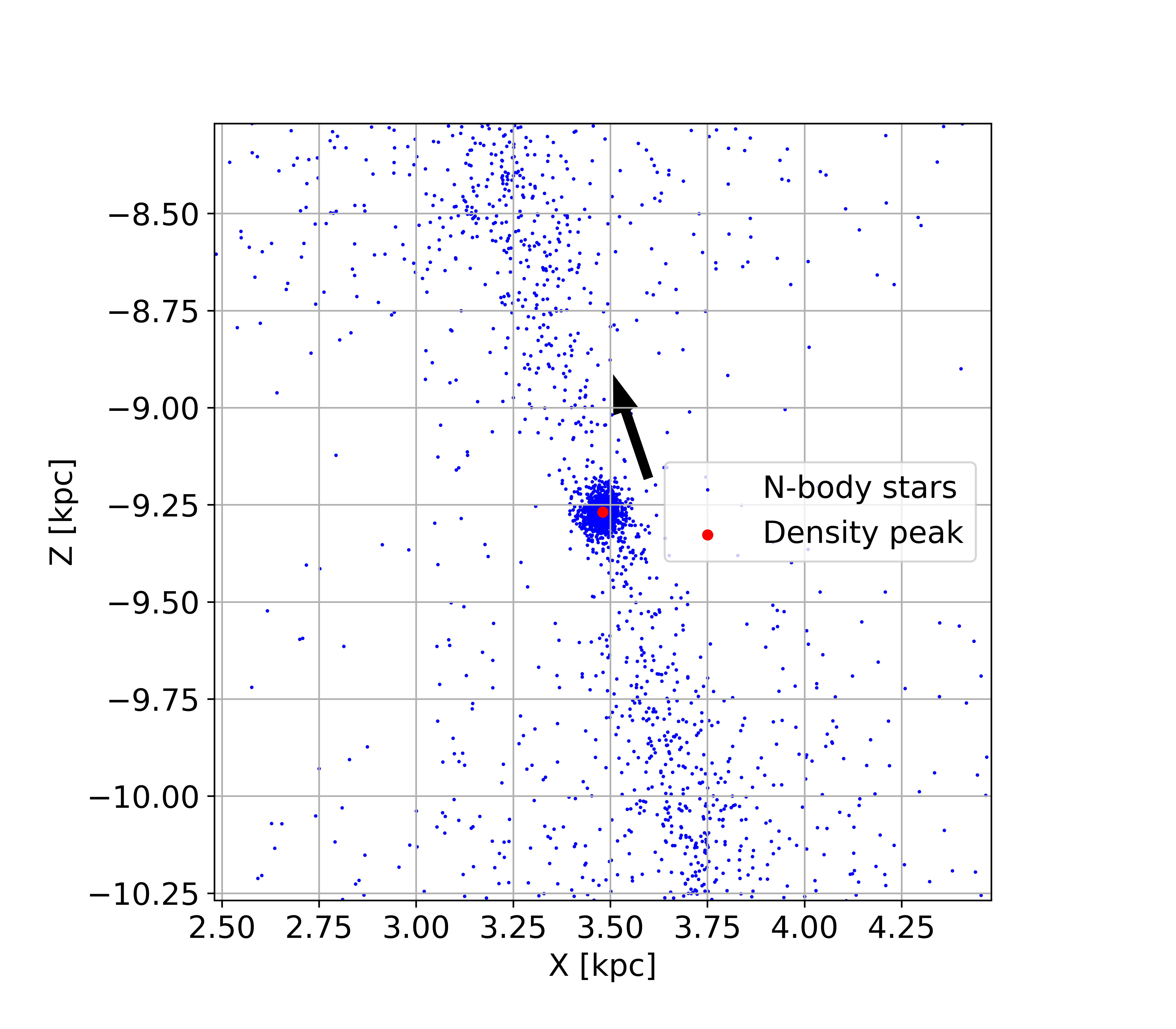}
      \includegraphics[width=0.7\columnwidth]{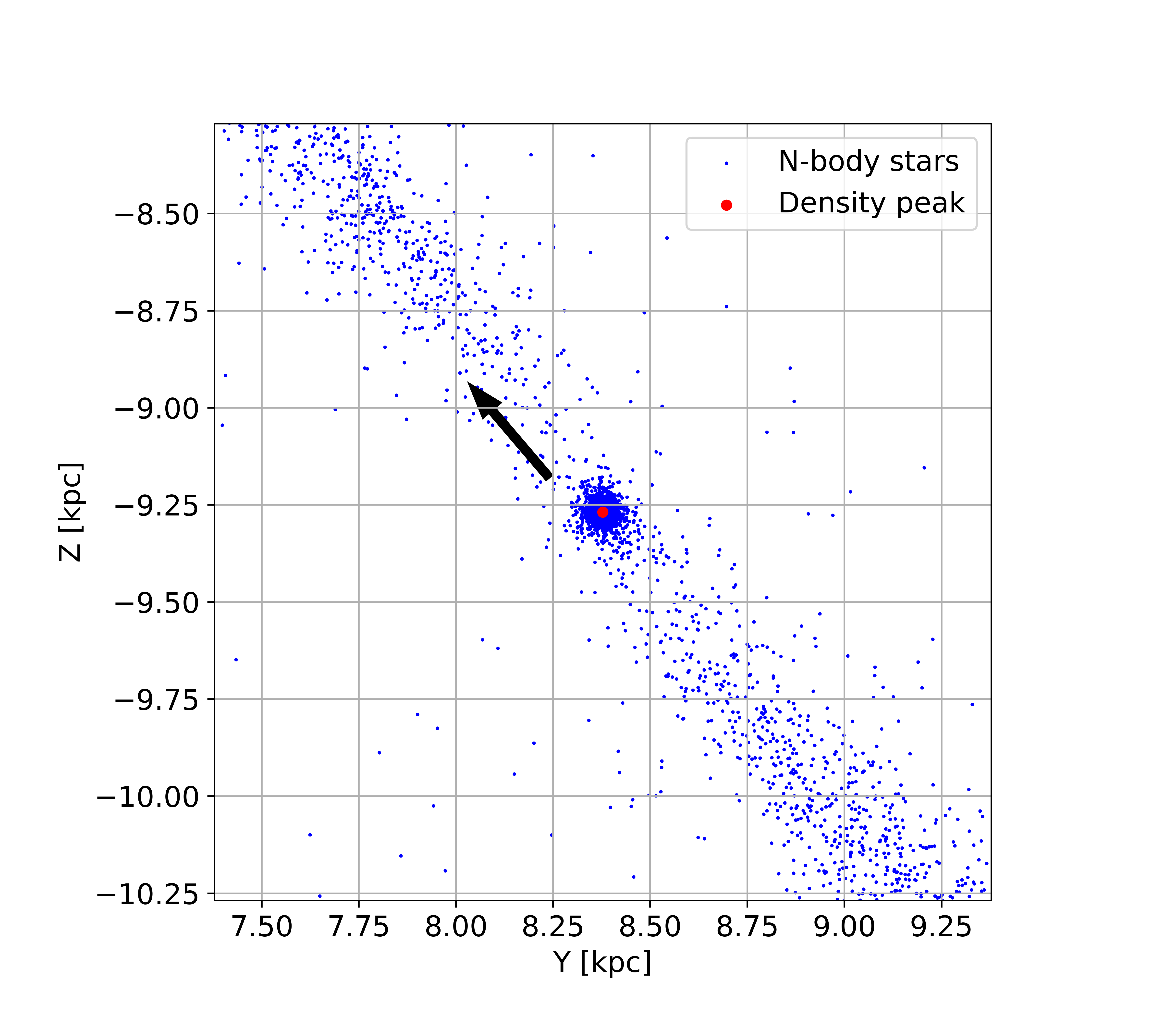}
  \caption{Present $N$-body snapshot (blue) in the $X-Y$, $X-Z$, and   $Y-Z$ planes. 
  The point of highest
  density of the $N$-body cluster is marked in red. The black arrow is parallel to the
  velocity of NGC\,6981 in projected plane.}
  \label{fig:fig8}
\end{figure*}

\begin{figure}
  \includegraphics[width=\columnwidth]{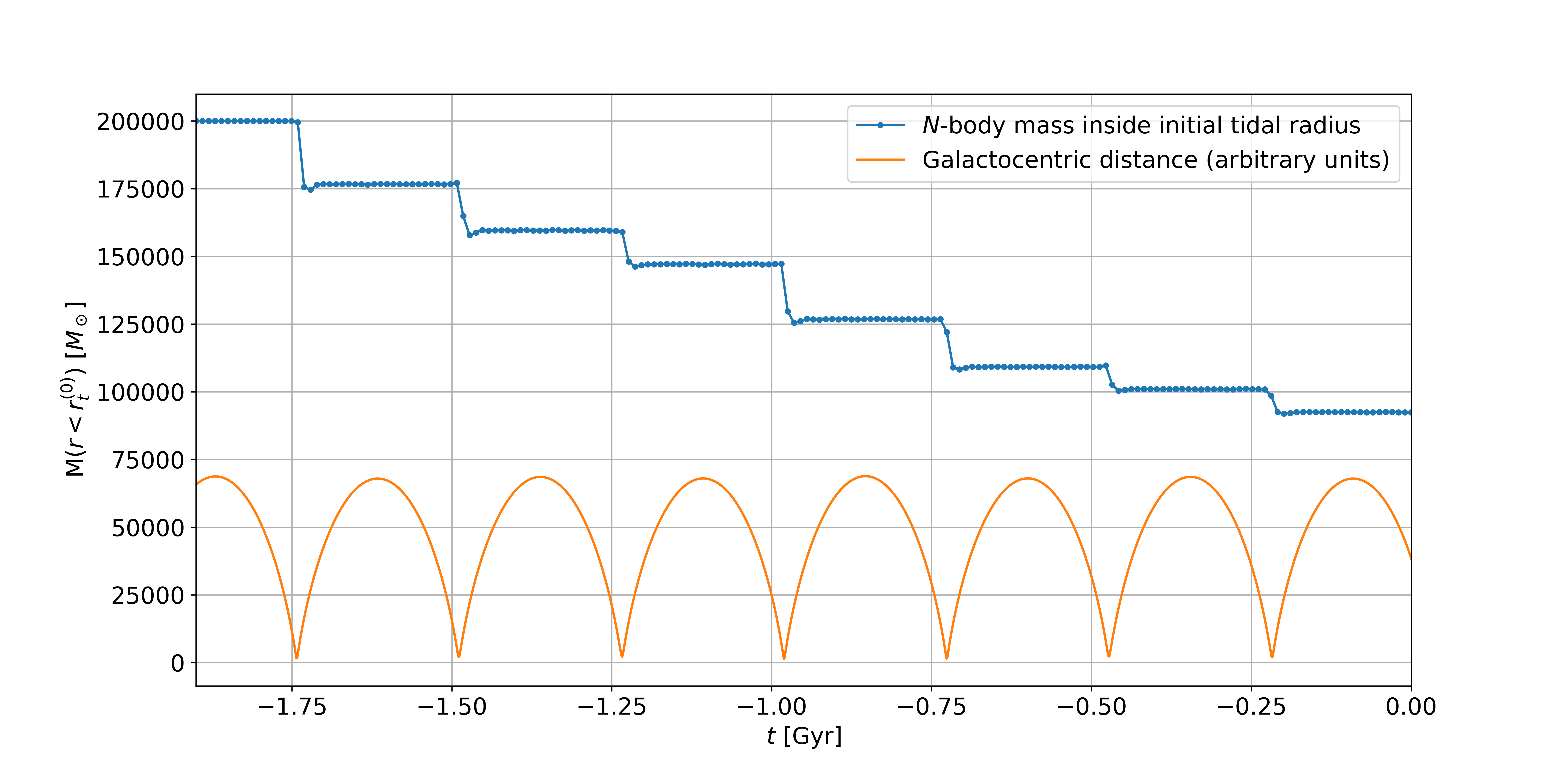}
  \caption{Mass evolution (blue dotted line) and Galactocentric distance in arbitrary units (solid orange line) of the $N$-body cluster.}
  \label{fig:fig9}
\end{figure}

Fig.~\ref{fig:fig7} shows that a large stellar stream with both leading and trailing tails has formed, 
which explores both disk and halo regions. The point of highest density in the present $N$-body 
cluster was approximated as follows. We started computing the center of mass of all the particles 
inside a box with edges $r_{\rm{box}}=50 r_c$, centered at the origin of coordinates. 
Then we computed a new center of mass by choosing the particles that lie inside a box with an 
edge 1 per cent smaller than the previous one, centered at the previous center of mass, and 
iterated this last step two hundred times. The resulting center of mass turned out to be a good 
approximation  to the density peak of the distribution. The regions close to the present $N$-body 
snapshot, projected onto the three coordinate planes are depicted in Fig.~\ref{fig:fig8}.
The offset of the present position of
NGC\,6981 with respect to the point of highest density of the $N$-body simulation is as expected: 
the backward orbit of the test particle initialized at the position of NGC\,6981 is followed forward
 approximately by the center of mass of the full set of $N$-body particles, which does not coincide with their peak density due to the streams. Fig.~\ref{fig:fig9} shows the mass of the $N$-body cluster inside a radius equal to the initial tidal radius $M(r<r_{t})$, as a function of time. 
 It is readily visible that the cluster loses stars only at its pericentric passages, each leakage during approximately 20 Myr.

\begin{figure*}
  \includegraphics[width=0.7\columnwidth]{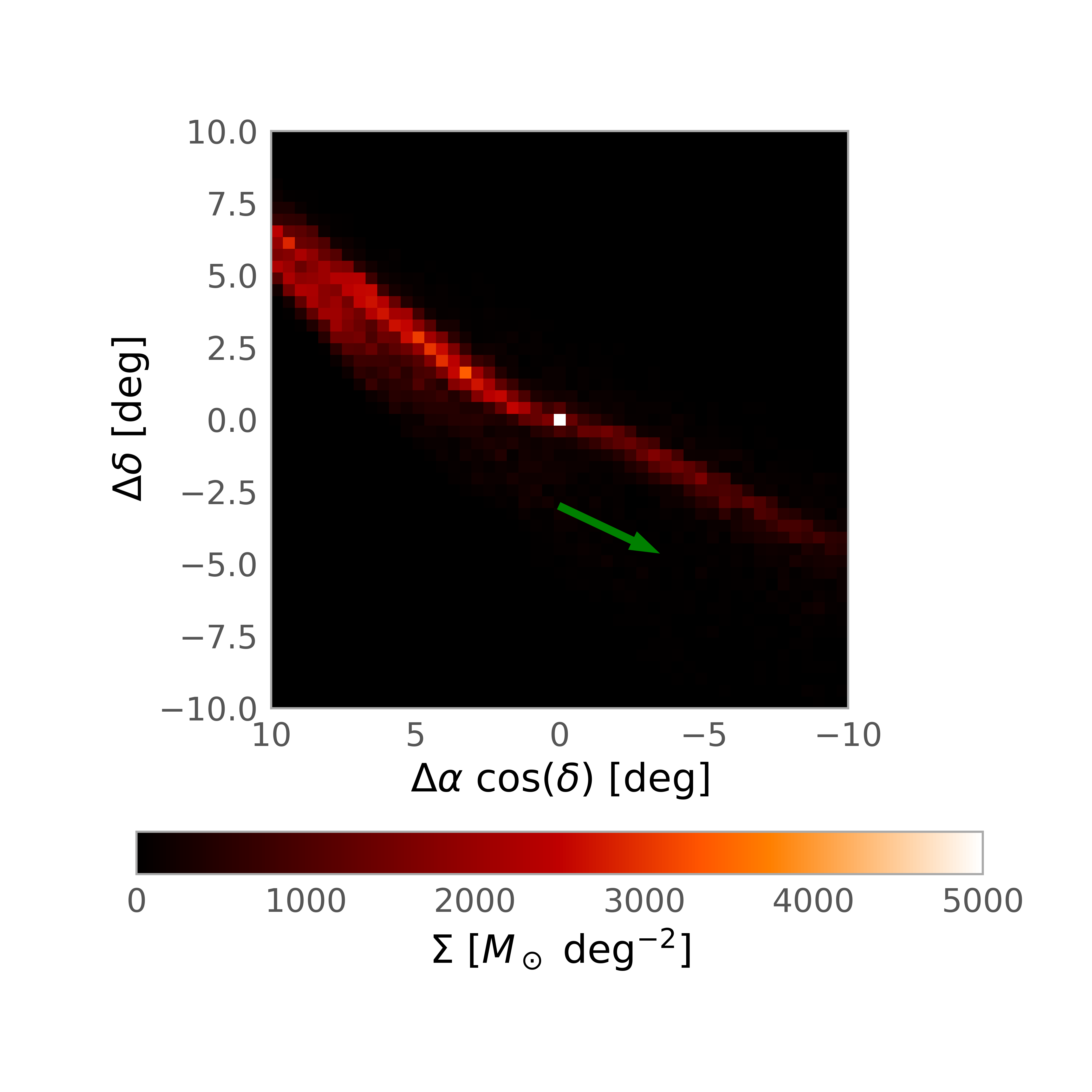}
  \includegraphics[width=0.7\columnwidth]{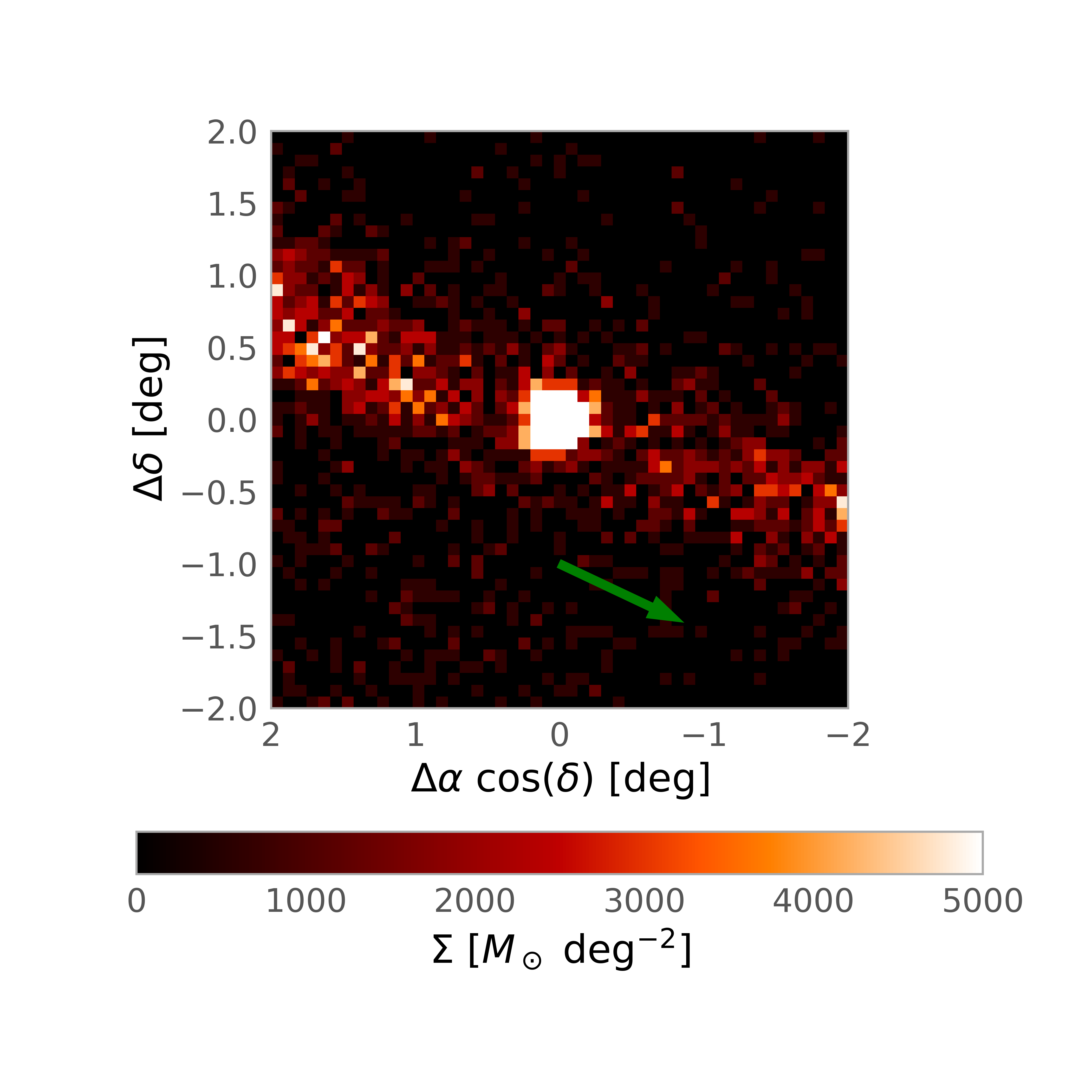}  
    \includegraphics[width=0.7\columnwidth]{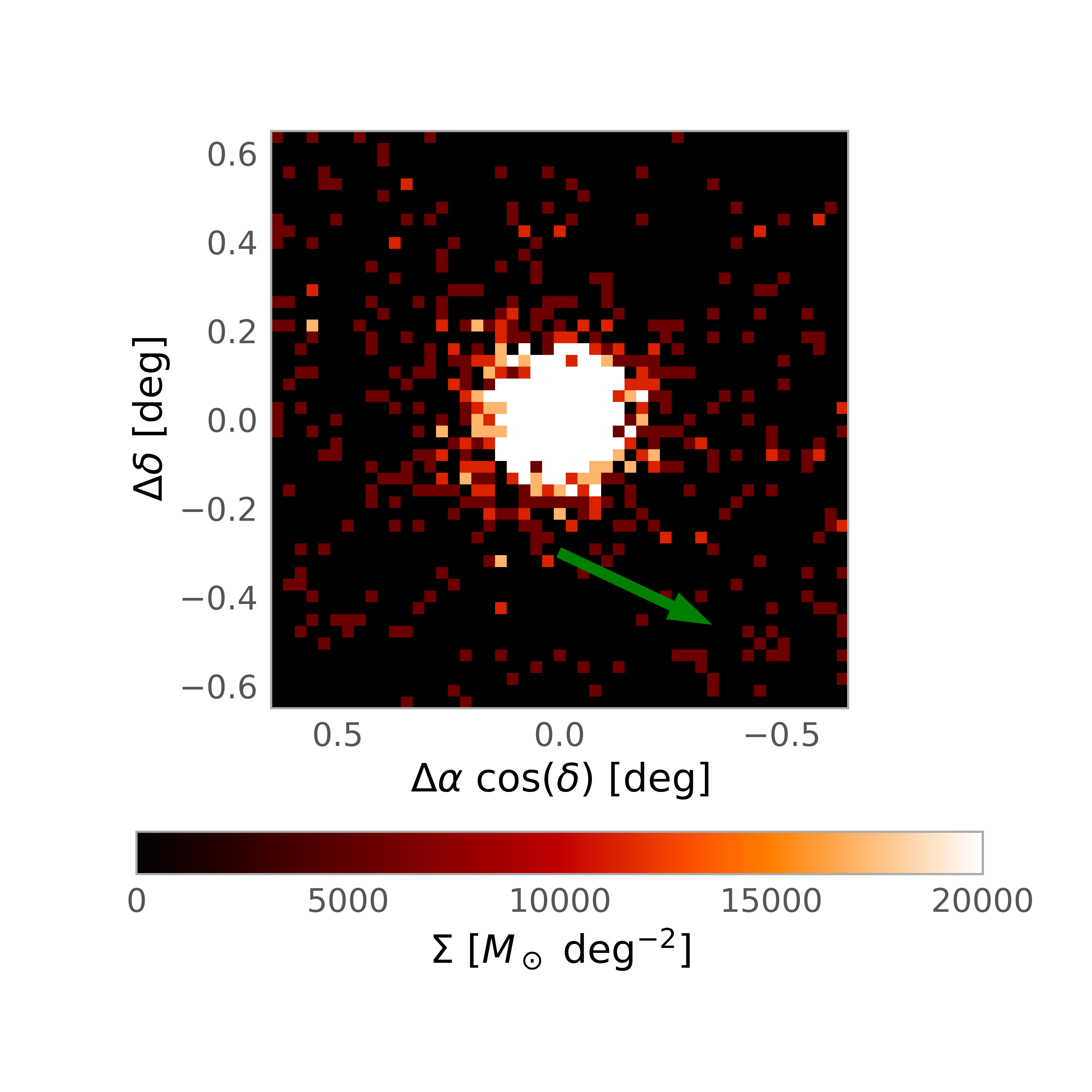}
  \caption{Surface density of the $N$-body cluster on the celestial sphere. The arrow is parallel to the measured proper motion vector corrected by the solar reflex motion with the 
\texttt{Gala} code \citep{pricewhelan:2017a}. The right panel can be directly compared to
Fig.~\ref{fig:fig6}.}
  \label{fig:fig10}
\end{figure*}

To compare the distribution of the $N$-body particles with the observed stellar density 
(see Fig.~\ref{fig:fig6}), we projected the former into the celestial sphere and
computed the surface mass density $\Sigma$($\Delta$((R.A.)cos(Dec.),$\Delta$(Dec.)), where 
$\Delta$(R.A.) and $\Delta$(Dec) are the R.A. and Dec. of a point of the sphere with respect to the position of the highest density peak of the $N$-body cluster.
In order to highlight the density variations, we have applied an upper threshold
value of $\Sigma_{\rm{thres}}=5000$ M$_\odot$ deg$^{-2}$; all the pixels above this
value were colored in white. 
Fig.~\ref{fig:fig10} shows that the tidal tails extend towards the North-East and the South-West directions, respectively.  Notwithstanding the clear streams in the larger fields, they seem to have disappeared in the smallest one. This is due to the fact that the stream  slowly drifts from the cluster, and the mass loss only happens  during very short intervals at the pericenter passages. Therefore, we expect a gap of streaming stars near the cluster, widening between pericenter passages.

\section{Conclusions}

We analyze the outer regions of the Milky Way globular cluster NGC\,6981, with the aim of
identifying extra-tidal features. The cluster caught our attention because it was associated to
the accreted Helmi streams, and deposited in an extremely inner halo orbit,
with a relatively high eccentricity and inclination angle with respect to the Milky Way plane  in the current orbital section. 
According to
\citet{piattietal2019b}, globular clusters with orbital parameters similar to those of 
NGC\,6981 have lost
relatively more mass by tidal disruption than globular clusters rotating in the Milky Way disk. 

We started by analyzing $BV$ public data sets, which revealed the presence of extra-tidal
debris distributed around the cluster.  Our DECam observations, which reach nearly 4 mag
below the cluster MS turnoff, confirmed our findings from a wider field of view. $BV$ and
DECam photometries were treated similarly. We relied on MS stars, particularly the
fainter ones,  because they are the first to
cross the Jacobi radius once they reach the cluster boundary driven by two-body relaxation.
Indeed, we found that the fainter the range of magnitudes considered, the more extended
the cluster stellar density map.

In order to monitor any differential change in the stellar density map caused by cluster stars
with distinct brightnesses, we split the long MS of the cluster into two segments of 
2 and 1.5 mag long, respectively, and we analyzed them separately. The magnitudes and
colors of the stars were corrected by individual interstellar reddening, which slightly varies 
across the field. 
Then, we statistically decontaminated the cluster MS segments from the presence of
field stars, and built stellar density maps using all the stars that remained unsubtracted after
the cleaning procedure.
We built their respective stellar density maps using a kernel estimator
technique, which revealed the presence of extra-tidal features oriented along
the opposite direction to the Milky Way center.

$N$-body simulations were finally performed to help us to understand the spatial
distribution of the extra-tidal debris. In our simulations, the cluster was represented by 
an  $N$-body system evolved during 2 Gyr in the adopted Milky Way potential. The disrupted stars
clearly form long trailing and leading tails, that are mostly parallel to the direction of the
cluster velocity vector, which is similar to the direction pointing to the Galaxy centre 
at the present time. 
We confirmed a decrease in the density
of extra-tidal stars near the cluster. Also, we found that
stronger extra-tidal features could be found by exploring larger areas around NGC\,6891.

\begin{acknowledgements}
M.F.M. and D.C. acknowledge support from the Universidad Nacional de La Plata (grant 11/G153).
M.F.M. thanks Juan Ignacio Rodriguez for maintaining the IALP server where
the $N$-body code was run.
M.D.M. thanks to the postdoctoral position CONICYT-GEMINI.
We thank the referee for the thorough reading of the manuscript and
timely suggestions to improve it. 

Based on observations at Cerro Tololo Inter-American Observatory, NSF’s NOIRLab (Prop. ID 2019B-1003; 
PI: Carballo-Bello), which is managed by the Association of Universities for Research in Astronomy (AURA)
under a cooperative agreement with the National Science Foundation.

This project used data obtained with the Dark Energy Camera (DECam), which was constructed by the 
Dark Energy Survey (DES) collaboration. Funding for the DES Projects has been provided by the US 
Department of Energy, the US National Science Foundation, the Ministry of Science and Education of Spain, 
the Science and Technology Facilities Council of the United Kingdom, the Higher Education Funding Council 
for England, the National Center for Supercomputing Applications at the University of Illinois at 
Urbana-Champaign, the Kavli Institute for Cosmological Physics at the University of Chicago, Center for 
Cosmology and Astro-Particle Physics at the Ohio State University, the Mitchell Institute for Fundamental 
Physics and Astronomy at Texas A\&M University, Financiadora de Estudos e Projetos, Funda\c{c}\~{a}o 
Carlos Chagas Filho de Amparo \`{a} Pesquisa do Estado do Rio de Janeiro, Conselho Nacional de 
Desenvolvimento Cient\'{\i}fico e Tecnol\'ogico and the Minist\'erio da Ci\^{e}ncia, Tecnologia e Inova\c{c}\~{a}o, the Deutsche Forschungsgemeinschaft and the Collaborating Institutions in the Dark Energy Survey.
The Collaborating Institutions are Argonne National Laboratory, the University of California at Santa Cruz, the University of Cambridge, Centro de Investigaciones En\'ergeticas, Medioambientales y Tecnol\'ogicas–Madrid, the University of Chicago, University College London, the DES-Brazil Consortium, the University of Edinburgh, the Eidgen\"{o}ssische Technische Hochschule (ETH) Z\"{u}rich, Fermi National Accelerator Laboratory, the University of Illinois at Urbana-Champaign, the Institut de Ci\`{e}ncies de l’Espai (IEEC/CSIC), the Institut de F\'{\i}sica d’Altes Energies, Lawrence Berkeley National Laboratory, the Ludwig-Maximilians Universit\"{a}t M\"{u}nchen and the associated Excellence Cluster Universe, the University of Michigan, NSF’s NOIRLab, the University of Nottingham, the Ohio State University, the OzDES Membership Consortium, the University of Pennsylvania, the University of Portsmouth, SLAC National Accelerator Laboratory, Stanford University, the University of Sussex, and Texas A\&M University.
\end{acknowledgements}



\end{document}